# Validation Requirements for AI-based Intervention-Evaluation in Aging and Longevity Research and Practice

Georg Fuellen[1,2,*], Anton Kulaga[1], Sebastian Lobentanzer[3,4], Maximilian Unfried[5,6], Roberto A. Avelar[1], Daniel Palmer[1], Brian K. Kennedy[5,6,7]*

*Corresponding Authors

[1]Institute for Biostatistics and Informatics in Medicine and Ageing Research, Rostock University Medical Center, Rostock, Germany
[2]UCD Conway Institute of Biomolecular and Biomedical Research, School of Medicine, University College Dublin, Dublin, Ireland
[3]Institute for Computational Biomedicine, Heidelberg University, Faculty of Medicine and Heidelberg University Hospital, Heidelberg, Germany
[4]European Bioinformatics Institute, Hinxton, Cambridgeshire, UK
[5]Healthy Longevity Translational Research Program, Yong Loo Lin School of Medicine, National University of Singapore, Singapore
[6]Department of Biochemistry, Yong Loo Lin School of Medicine, National University of Singapore, Singapore
[7]Department of Physiology, Yong Loo Lin School of Medicine, National University of Singapore, Singapore

**Abstract.** The field of aging and longevity research is overwhelmed by vast amounts of data, calling for the use of Artificial Intelligence (AI), including Large Language Models (LLMs), for the evaluation of geroprotective interventions. Such evaluations should be correct, useful, comprehensive, explainable, and they should consider causality, interdisciplinarity, adherence to standards, longitudinal data and known aging biology. In particular, comprehensive analyses should go beyond comparing data based on canonical biomedical databases, suggesting the use of AI to interpret changes in biomarkers and outcomes. Our requirements motivate the use of LLMs with Knowledge Graphs and dedicated workflows employing, e.g., Retrieval-Augmented Generation. While naive trust in the responses of AI tools can cause harm, adding our requirements to LLM queries can improve response quality, calling for benchmarking efforts and justifying the informed use of LLMs for advice on longevity interventions.

**Introduction**

Interventions in longevity research and practice range from genetics, pharmaceutical compounds and supplements to diet, exercise and stress management. A major goal is to determine whether such "geroscience" interventions are effective and safe, and why. Such "intervention analytics" need to tailor interventions to people based on biomarkers, revealing whether an intervention may trigger certain desired changes of biomarkers, and how [1]. Analyzing interventions also includes their (re-) positioning on the population level. In all cases, insight into the effectiveness of interventions will scale with the comprehensiveness of the data collected and the sophistication of the analytics employed.

In high quality studies, in-depth clinical lab measurements, multi-omics, and consideration of as many (socio-)environmental factors as possible are critically important [1]. Furthermore, enlarging the breadth and depth of data analyses and thereby considering existing knowledge in a detailed manner must be equally emphasized [2]. Currently, intervention data is often discussed and analyzed by referring to canonical databases, such as the Gene Ontology (GO) or KEGG in case of gene expression analyses [3]. Such "classical" bioinformatics is certainly insightful and will continue to provide a frame of reference; however, the use of AI should make it possible to connect the changes induced by an intervention with a *wide array of knowledge* in a *much more detailed* manner. After all, while databases such as GO aim for some comprehensiveness, they cannot reflect all the intricate biological interrelationships known in current molecular biology. Thus, AI may improve upon classical evaluations based on literature appraisal, bioinformatics and biostatistics. (Here, AI specifically refers to LLMs, which are "a type of AI model using deep neural networks to learn the relationships between words in natural language, using large datasets of text to train" [4] and which, on that basis, generate new text in response to text queries; current examples are the GPT and Claude families of models.)

As an example, we may find out by gene set enrichment analyses that some genes affected in a cell experiment are surprisingly often known to be involved in "apoptosis", and predict that these cells are struggling with cell death. *However,* connecting the dots from many papers and/or database entries may show that the majority of these genes are known to regulate a specific subset of the apoptosis pathway, potentially in specialized cell types. This regulation may frequently be observed for senescent cells, as opposed to healthy cells, suggesting that interventions known to affect these genes may be senolytic. A case in point are the genes involved in "senescent cell anti-apoptotic pathways" (SCAPs) [5]. An intervention specifically downregulating these genes might facilitate elimination of senescent cells by a known mechanism, that is, by attacking the SCAPs. Based on similar examples, much-needed benchmarking datasets for the comprehensive analysis of interventions, e.g. by AI, need to be curated, considering detailed molecular (omics) data, but also diverse datasets including complete blood count analyses, data from wearables, questionnaire responses, and others.

Here, we outline how AI, particularly LLMs and knowledge graphs (KGs), can aid with analysis of the increasing amounts of data in aging and longevity research. Such analyses are expected to yield improved *biomarkers of aging* [1][6], as well as more accurate and faster *evaluations of geroscience interventions* with respect to efficacy and safety [2]. We have listed a set of requirements for good evaluations, alongside some best practice suggestions. Importantly, evaluations must be benchmarked, based on curated evaluations adhering to the requirements we have listed. Ultimately, on the population level, the added value of AI can only be proven by systematically studying positively evaluated interventions prospectively in humans. However, better AI-based evaluations can (1) prioritize which interventions should be tested in expensive clinical studies, (2) aid in designing the studies to optimize the likelihood of successful outcomes and (3) potentially suggest non-classical clinical trial designs targeted toward community dwelling, largely healthy individuals. In the *Discussion*, we list the strengths and limitations of our work, and compare it to similar work outside of geroscience.

**Key requirements for using AI methods to evaluate geroscience interventions**

We propose a set of eight requirements in Table 1: three basic, three intervention-specific, and two geroscience-specific. For each of them, we suggest best practices given the current state of the art. On one hand, we wish to evaluate geroscience interventions in general, i.e. at the population level. On the other hand, we want to give advice to individuals for such interventions. The query/question then is the description of a biomedical intervention together with some biomarker values defining the context of its application (see also Fig. 1a for an example), and the desired answer is an evaluation of that intervention, telling us whether it is expected to foster healthspan and lifespan, and why.

**Table 1. Requirements for evaluating geroscience interventions, by AI tools in particular, and consequences for the choice of method and best practice.**

|   | Requirement | Choice of Method / Best practice |
|---|---|---|
| 1 | *Correctness* of the evaluation results. Input data quality checks. | *Use of methods and tools with a high probability of correctness, optimally* based on community benchmarks. *For example,* combining LLMs with KGs [7]. Specific directives/prompts to check input data quality. |
| 2 | *Usefulness* and *comprehensiveness.* | *Use of methods and tools such as LLMs with Retrieval-Augmented Generation mechanisms* [7], *and large KGs* in synergy with LLMs. Integration of auxiliary data analyses, e.g., for quantitative data [8]. |
| 3 | *Interpretability* and *explainability* of the evaluation results. Clarity and conciseness of the results and the given explanations. | Use of interpretable/explainable methods and tools, *e.g., explanation and visualization of how the method calculated the answer; explicit explanation of the answer by the method* or by auxiliary tools [9]. |
| 4 | Specific consideration of *causal mechanisms* affected by the intervention. | *Specific directives/prompts, or use of causal analysis* [10], *or of tools elucidating underlying biological mechanisms.* Augmentation by retrieval and analysis of relevant randomized controlled trial (RCT) data. |
| 5 | Consideration of data in *a wide ("holistic") context:* (a) efficacy *and* toxicity, and evidence for the existence of a large therapeutic window; (b) analyses in an "interdisciplinary" setting. | *Specific directives/prompts;* use of knowledge, *e.g.,* from toxicology. Specific directives/prompts to consider an "interdisciplinary" setting, *e.g.,* socio-environmental factors, economic factors, and sustainability. |
| 6 | Enabling *reproducibility, standardization*, and *harmonization* of the analyses. | *Specific directives/prompts* to follow standard procedures of analysis, and transparent documentation of all analysis steps [1]. |
| 7 | Specific emphasis on *diverse longitudinal large-scale data.* | *Augmentation by data retrieval and analysis regarding large-scale longitudinal data* [1,6]. |
| 8 | Specific emphasis on results that relate to *known mechanisms of aging.* | *Specific directives/prompts* for analyses that highlight or prefer known mechanisms of aging based on consensus listings [11,12]. |

Requirements 1-3 demand that analyses and evaluation results are correct, useful and comprehensive, and that they are explained with detailed rationales. Here, AI features specific advantages by enabling comprehensive analyses, and it encounters specific challenges in returning correct and explainable results. Requirements 4-6 aim to enforce causality, consideration of a wide range of factors related to the applicability and consequences of the intervention (efficacy/toxicity, and interdisciplinary consideration of psychological and social factors), and alignment of the analyses with standard operating and reporting procedures. Finally, requirements 7+8 focus on longitudinal data, as these provide advantages compared to cross-sectional data, and on analysis results relating closely to known mechanisms of aging [11] [12].

In Fig. 1, we exemplify how consideration of the requirements *as a whole* can improve the evaluation of an intervention in practice. As an example we take rapamycin, a prescription drug frequently used off-label with the idea to promote healthspan and lifespan [13] [14]. We query the LLM about the value of taking rapamycin in some specific biomarker context, including age and weight data (Fig. 1a), and we note that it responds with more relevant details and caveats if the question is appended by an outline of the requirements of Table 1 / Fig. 1b, that is, "*Please fulfill the following criteria in your response: correctness, usefulness and comprehensiveness, human interpretability, as well as consideration of causality, toxicity and holistic/interdisciplinary evidence, standardized ways of analysis and reporting, longitudinal data and known aging biology*" (Fig. 1c). The assessment of the output by another model, checked by us for plausibility, is also improving (Fig. 1d). Notably, the positive effect of appending the requirements was largest for GTP4 as assessed by Claude Opus, and for consideration of data in a wide context, including toxicity in particular. To broaden our test cases, and since many interventions are not drugs, but rather supplements or even specific dietary items, we also considered strawberries and quercetin (Supplementary Figures 1-4). Similar to rapamycin, we find that our proposed listed requirements have a positive effect at meeting key criteria. Therefore, the applicability of our proposed paradigm appears to be robust across different kinds of interventions. Full recordings are available (Supplementary Text 1).

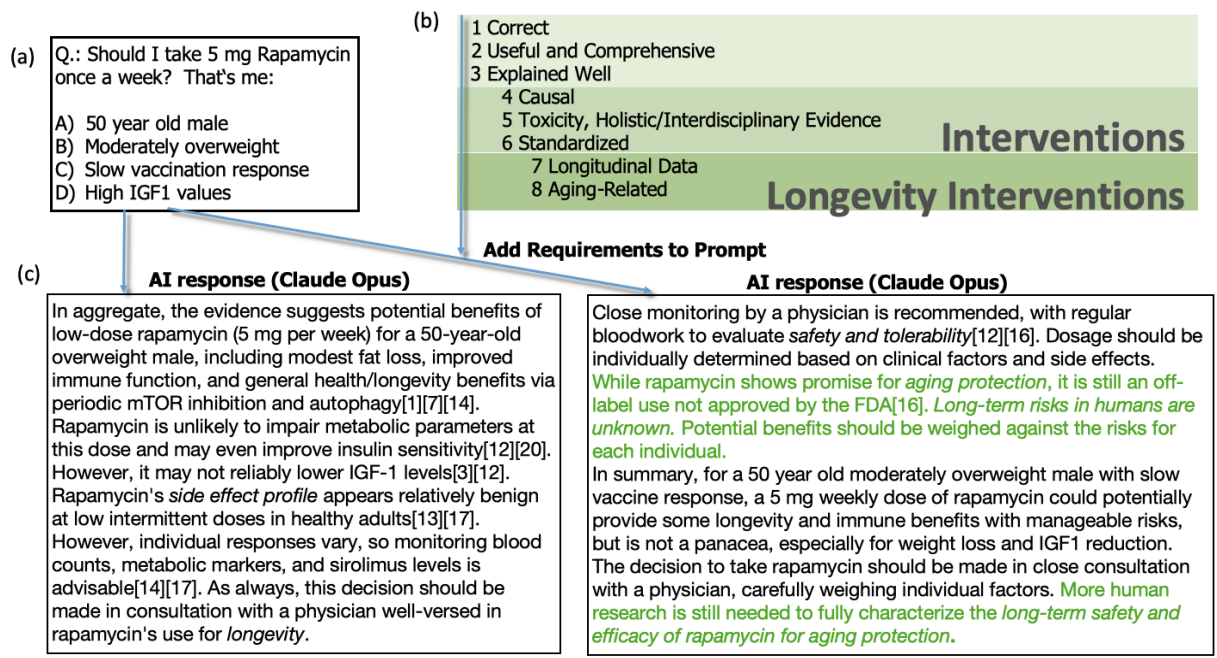

Symbols: √, good; (√), moderate; x, bad. Perplexity.ai Pro was used in April 2024 to ask Claude 3 Opus and GPT4 Turbo, starting a new thread each time. Evaluations were done by GPT4 in case of Claude output, and by Claude in case of GPT4 output. Improvements were marked in green. From the answers, only the summaries are shown in (c).

**Fig. 1: (a)** Sample query to LLMs about intervening with rapamycin in a specific biomarker context. **(b)** Schema of requirements for evaluating biomedical interventions. **(c)** Sample LLM answers, from Claude Opus. The green text highlights improvements due to explicitly listing all requirements as part of the query prompt; formatting in italics was added by us. **(d)** Adherence of an LLM (GPT4 / Claude) to the requirements, assessed by the "competitor" LLM (Claude / GPT4).

**Implementing the requirements: Importance and judgment criteria**

*Evaluating interventions, use cases and scenarios.* We can frame intervention evaluation as a question-and-answer process. For an individual or a population, an evaluation is requested for a specific intervention, usually providing some context, e.g., biomarker values. (In Supplementary Text 2, we define precisely the terms "intervention" and "intervention evaluation", matching the intuitive meaning of everyday language). For simplicity, here we focus on the case of an individual asking "Should I take 5 mg rapamycin once a week?" together with some simple biomarker data (Fig. 1). An evaluator (a human operating a traditional analysis pipeline, or an AI system) responds. For a meaningful response, the evaluator needs access to study data that shows how individuals (potentially including animals) with the same or similar biomarkers have responded to the same or similar interventions in the past. Matchings between the questioner's biomarker data and biomarker data from past studies serve as evidence for the evaluation, which could be positive if the matchings suggest potential health improvements. Judging the quality of the evaluation is then done based on the requirements of Table 1, which we simulated by the assessments in Fig. 1d.

**Correctness**. We will first discuss correctness (also referred to as factuality [15]), which may be considered particularly important for geroscience interventions, as these are to be implemented by healthy people, with the goal of staying healthy longer. It is also the hardest to define precisely and to check rigorously. *In rare cases*, knowledge about the query intervention is available and can be matched directly by the evaluator. In the rapamycin example, the 5 mg per week dose may already have been shown to be effective for individuals with the same biomarkers as the questioner. Then again, findings of some past intervention studies may not be generalizable because the context of the findings may be important. For example, there is always the risk of cohort effects; results from the past may not hold because individuals today live in a different environment, etc. *More frequently*, the evaluator can only match the query information with what is known based on performing some advanced reasoning. For example, it may have to assume that (1) biomarkers are comparable because of their similar response to certain kinds of interventions (such as some sets of inflammatory cytokines), or (2) that interventions are similar (such as various fasting regimes), or (3) that dose-dependent intervention effects can be extrapolated. Then, both the knowledge as well as the reasoning can be a source of error for human and AI-based evaluators alike. Pragmatically, both are judged based on past performance, that is, by benchmarking (see also below). Furthermore, quality checks should be implemented, *at least* by checking ambiguity of the input describing the intervention and its context, and by investigating the plausibility [16] of the input data, of the reasoning and of the resulting output. If some sources of knowledge were not considered in some evaluation steps, comparing preliminary results with this knowledge is useful and important, although whenever possible, the totality of all the knowledge must be used for intervention evaluation.

**Usefulness and comprehensiveness.** Irrelevant (but true) information and missing (but true and relevant) information naturally diminish the quality of an evaluation. Moreover, to be comprehensive, we must not only consider the data but also the analysis and the reasoning steps [17]. Aging is a process affecting all aspects of life, so we must consider a wide range of connections with already established knowledge, which may be findable in principle by in-depth searches of the literature and in-depth comparison with other molecular data, as deposited in public databases. A case in point is the apoptosis example from the *Introduction*. The knowledge about SCAPs in the literature and in databases could be represented as (part of) a detailed KG, with genes, proteins, metabolites, and other entities as nodes and their interrelationships as edges, optimally including cellular context. LLM workflows with Retrieval-Augmented Generation (RAG) could then retrieve and consider all relevant papers available, as well as the KG data on SCAPs and related processes, for a comprehensive analysis.

Usefulness considers that the evaluation must be relevant, clear, concise, and tailored to the user's needs. Thus, appropriate user feedback, while subjective, is the gold standard. To be comprehensive, using and reporting all the evidence pertinent to the evaluation of the intervention in a totality-of-the-evidence approach is recommended, as it is closest to employing high-quality meta-analyses. Human meta-analysis efforts are painstaking and slow, and AI may be able to optimize the effort of correctly aggregating knowledge into meta-analyses. Then again, human meta-analyses are good benchmarks if they are not yet part of any model training data.

***Explainability.*** Explainability/interpretability refer to a major field of AI research, known as explainable AI (XAI) [18] [19]; different notions of interpretability were recently defined, with healthcare in mind [20]. In particular, we may seek explanations in an "intrinsic" fashion, by explaining how the answer was calculated by the method itself. In principle, all the analyses and the reasoning of the evaluator of an intervention may be recorded, analyzed, summarized and visualized for better human understanding. Moreover, an aspect of "intrinsic" explainability for AI is that an explanation can be explicitly requested from the evaluator. Last but not least, and specifically important for evaluating geroscience interventions, the intrinsic explanation of the evaluation results, and the evaluator's explanation for these (and for the intermediate reasoning steps) should, whenever possible, refer to the biomolecular mechanisms already known to be triggered by the intervention. The alternative to an intrinsic explanation is to extrinsically derive the explanation in a post-hoc fashion, investigating input-output relationships [9]. Sometimes a trade-off between explainability and correctness/accuracy is noted [21], but retrieving explanations can also improve correctness [9]. Moreover, for human interventions, explainability is crucial because of adherence and regulatory issues.

Explanations for an evaluation often depend on the method used to generate it. Unsupervised (or supervised) machine learning, in the simplest case by using principal component analysis (PCA), can provide a visualization of the data in a similarity space. More generally, a plane or a 3-dimensional space can be used to visualize both the input intervention and the interventions to which it is compared. This space may be a projection of the "embedding", "representation", or "latent" space employed by the evaluator. Further, and not unique to PCA, proximity between interventions in similarity space can then be an important part of the explanation for an evaluation, allowing to refer to interventions similar to the one under consideration. In a plausible similarity space, interventions with similar effect and toxicity profiles should cluster together. We are mostly concerned with explanations that are "local", that is, calculated for a certain input. In this case, specific benchmarks may be curated based on case studies of explanations for evaluations of known geroscience interventions. More generally, and for "global" explanations of the general workings of an evaluator, more generic XAI benchmarking data may be used [18].

***Causality, interdisciplinarity, standards, and specifics of geroscience intervention evaluation.*** Regarding the advantages of considering causality, we consider the rapamycin example. If a high-quality RCT demonstrates that rapamycin triggers health benefits, it could be recommended to an individual as long as the biomarker context matches sufficiently well. In contrast, observational studies may face confounding variables; for instance, health-conscious individuals taking rapamycin and engaging in other health-seeking behaviors, complicating causal links to rapamycin's effects [13]. Moreover, any biomedical intervention should be evaluated by explicitly searching for, analyzing and considering toxic ("adverse") side effects, as well as intervention effects that may relate to the "social hallmarks of aging" [22] and other interdisciplinary considerations. Furthermore, all evaluations should be conducted according to standard operating procedures (part of which may be inspired by this manuscript). Standards in analyzing and reporting also enable better reproducibility [1]. On the meta-level, standards for reporting trials of decision support systems driven by AI and for systematic reviews and meta-analyses on AI in healthcare are available or in preparation [23] [24]. Finally, we suggest that interventions into the aging process are naturally best described by time series, that is, longitudinal data, and that the knowledge about aging processes accumulated until now is sufficiently correct, useful and comprehensive so that evidence matching this knowledge is more likely to be correct, compared to evidence that does not align with current knowledge [1].

Establishing causality usually requires a specific kind of ground truth typically based on intervention data, optimally RCTs, but also mechanistic studies, e.g., in animals or in-vitro; these may be used for benchmarking. Benchmarking based on human curation is also the current gold standard for evaluating the LLM's consideration of toxicity and other adverse outcome data, of non-molecular interdisciplinary data, and the use of standards, as well as the use of longitudinal data and consensus knowledge about aging-associated processes. The latter task is quite difficult because it can be challenging to explain the longevity benefits of interventions mechanistically.

**Evaluation scenarios, and AI method choice for evaluating interventions in geroscience**

*Evaluation scenarios.* As described, an intervention may be evaluated for an individual or a population, in some context described by biomarker data (age, weight, laboratory values, omics, etc.) that should be provided together with the query. These biomarker data may or may not be differential, that is, available for two or more points in time. Given minimal context, general advice can be given; some interventions, such as physical activity, albeit at different types and intensities, can be recommended in almost any context. If more biomarker data are provided for a single point in time, the advice can be more specific. Some interventions are known to improve health for people in a specific age range, or with specific blood count measurements. In the ideal case, *changes* of biomarkers in time are provided as context, and the advice can be personalized by comparing these changes to known intervention-triggered changes, aiming for a reversal. Such a reversal on the population level, considering population-level changes as the aggregates of changes on the individual level, may permit the repositioning of an intervention, checking whether it reverses changes known to reduce health and survival, without being associated with unwanted effects. For all these comparisons, further context of the changes must be considered (e.g., species, in-vitro vs in-vivo, …), aiming for maximum similarity on as many aspects of the analysis as possible. To address comprehensiveness and our other requirements in the best possible way, empowering LLMs is an important way forward.

*Counteracting LLM-specific limitations via Retrieval-Augmented Generation (RAG).* To maximize correctness and comprehensiveness of LLM responses, the current state-of-the-art suggests the application of prompt engineering via RAG mechanisms in particular. This includes the integration of structured and unstructured knowledge repositories, and the chaining of models for executing arbitrary code [7]. Structured data are represented in KGs, which can be queried by an LLM [25,26,27,28]. Unstructured data (e.g., scientific literature) can be made accessible to semantic similarity search by embedding the texts in a vector database [7,29]. Data is then retrieved based on the user's query and used to supplement the prompt in order to enable more correct responses by the LLM. Self-correcting prompt strategies [30,31] as well as multiagent debate methods [32] augment correction of some of the hallucinations and logical mistakes. In addition, many current-generation LLMs can write and parametrize arbitrary code, complementing their language abilities by providing a robust (and human-assessable) mathematical computation [33]. In the simplest cases, this amounts to calculating a mean or regression coefficient given structured numerical data. However, given an API (Application Programming Interface), LLMs can parameterize arbitrarily complex software to perform in-silico experiments with respect to a user's question [34]. For instance, a digital twin could be interrogated for interventional effects, or a genomics foundation model could be queried for the impact of a genetic perturbation. The results of these secondary models could then be returned to inform the primary LLM's response.

**Discussion and Conclusions**

We aimed for a complete list of relevant requirements that intervention evaluations should fulfill (Table 1). Some of these (1-3) are valid for any result of a scientific investigation, while some others (7+8) were specific for evaluating geroscience interventions. Some of our suggestions for method choice were also generic for most kinds of analysis, while others were specifically applicable to methods based on AI, and on LLMs (and KGs) in particular. A similar list was recently compiled, with an exclusive focus on LLMs in healthcare, but not considering intervention evaluation in any particular way [4]. There, *recency* was listed in lieu of comprehensiveness, also suggesting RAG for mitigation. *Accuracy* and *coherence* were listed, closely matching correctness, and mitigation strategies were given in detail, including better training, prompting and fine-tuning, while we refer to the use of benchmarked tools, and KGs (which foster the desired "true semantic knowledge"). *Transparency and interpretability* were listed as well, suggesting detailed responses with justification for the references cited. For evaluating LLMs on medical evidence summarization, factual consistency and comprehensiveness were suggested recently [35] and they align with our setup, alongside coherence (fluency) and harmlessness (no misinformation). Another recent and extensive review considered the effectiveness of healthcare conversations [36]; various flavors of correctness ("groundedness"), comprehensiveness, usefulness and robustness were noted.

There are also some requirements not listed in Table 1 because they are self-evident. For example, for any choice of method, its needs in terms of (computing) resources, and its availability (in terms of code and tutorials) are of course of high relevance; for some use cases, privacy and security may also be specifically important. Also, we did not cover some requirements closely related to the ones we are listing. For example, "reliability" was described and discussed [37], which is closely related to correctness but places an emphasis on robustness in the face of uncertainty. Also, some more limitations of LLMs related to a lack of correctness, including dependency on specific prompting and external factors, and biases, were considered [38] [their figure 3]. Finally, we did not explicitly consider the alignment of LLMs with human values [39], which is, however, implicitly covered by correctness, usefulness, and interdisciplinary consideration of, e.g., social factors (our criteria 1, 2 and 5).

Primarily for simultaneously addressing the core requirements of correctness, usefulness and comprehensiveness, we are lacking benchmark datasets to estimate the added value of AI that we believe to exist. Judging LLM-based assessments by LLMs, as done by us (Fig. 1d), is subject to all the known limitations of LLMs, and it is inherently circular. Simple curated benchmarks can be generated along the lines of the examples we presented in this manuscript. More sophisticated benchmarks shall consider the evaluation of an intervention in its entirety; these amount to meta-analyses. In an evaluation using AI models, benchmark contamination must be specifically considered [40]. The ultimate proof-of-concept must however be done in a prospective manner, showing that AI-based evaluations are clearly predictive of success in human studies, that is, in clinical trials.

Our outline for judging the AI-enhanced evaluation of (geroscience-based) interventions tried to provide a snapshot of the intersection of two fast-evolving fields, where we can expect a lot of progress in upcoming years. The ultimate goal could be a jack-of-all-trades tool, potentially designed and implemented by AI, or at least put together with the help of AI. To match new and old data in optimal ways, this tool would necessarily feature an all-encompassing integrated knowledge of biology; it would include features of a universal biomedical simulator [41]. This simulation could then be personalized to a specific human being, i.e. it would represent a digital twin [42]. Then again, human checking of these simulations and their results on some abstract level will, at least for a long time, provide a safety net against incorrect (and potentially harmful) evaluations [4].

Please find the supplement at
https://drive.google.com/file/d/1xHKqNWs4wvgclnE_jODvDKrhtDEyncPU/view?usp=sharing